\documentclass[12pt]{JHEP3}

\pdfoutput=1
\usepackage{bm}
\usepackage{epsfig}
\usepackage{citesort}
\usepackage{graphicx}
\usepackage{multirow}

\def\be{\begin{equation}}
  \def\ee{\end{equation}}
\def\bea{\begin{eqnarray}}
  \def\eea{\end{eqnarray}}

\title{Crossing Statistic: Reconstructing the Expansion History of the Universe} 

\author{Arman Shafieloo\\
	Institute for the Early Universe, Ewha Womans University\\ 
	Seoul, 120-750, South Korea\\
	E-mail: \email{arman@ewha.ac.kr}}

     
\abstract{We present that by combining Crossing Statistic \cite{Crossing,Crossing_B} and Smoothing method~\cite{Shafieloo06,Shafieloo07,Shafieloo10} one can reconstruct the expansion history of the universe with a very high precision without considering any prior on the cosmological quantities such as the equation of state of dark energy. We show that the presented method performs very well in reconstruction of the expansion history of the universe independent of the underlying models and it works well even for non-trivial dark energy models with fast or slow changes in the equation of state of dark energy. Accuracy of the reconstructed quantities along with independence of the method to any prior or assumption gives the proposed method advantages to the other non-parametric methods proposed before in the literature. Applying on the Union 2.1 supernovae combined with WiggleZ BAO data we present the reconstructed results and test the consistency of the two data sets in a model independent manner. Results show that latest available supernovae and BAO data are in good agreement with each other and spatially flat $\Lambda$CDM model is in concordance with the current data.  

}

\begin{document}

\section{Introduction}                        
\label{sec:introduction}

Reconstructing the expansion history of the universe and properties of dark energy is one of the main goals of today's cosmology to understand our universe and its components. There have been many approaches in last decade proposed to do the reconstruction of the expansion history and one can generalize them in two categories of parametric and non-parametric methods. Parametric methods are viable approaches if we know the actual class-form of the phenomena we are studying and we can use them to put constraints on the parameters of the model. See~\cite{davis07,serra07,eric08,sollerman09,kilbinger10} for details of data analysis and methods of parametric reconstruction of the properties of dark energy using supernovae data. However dealing with phenomena that we have no clear idea about its nature and behavior, using parametric methods can be misleading since the underlying actual model might not be covered by the assumed parametric form. Dealing with uncertainties in the dispersion of the data adds another complication to the analysis and leave us with no clear way to find this fact that we might have chosen an inappropriate parametric form. This raises the importance of the non-parametric and model independent approaches to find out the behavior of the phenomena in a more direct way by avoiding parametrizing cosmological quantities~\cite{daly03,huterer03,wang04,wang05,Shafieloo06,Shafieloo07,sahni08,zunckel08,shafieloo09,clarkson10,Shafieloo10,nesseris10,holsclaw10,crittenden11}. However non-parametric approaches have their own shortcomings. For instance, estimation of the errors can be a tricky task in many cases since in some methods one cannot easily assign the degree of freedom in the likelihood analysis. For a review over this subject look at~\cite{review_DE}.    

Smoothing method was proposed to reconstruct the expansion history of the universe in a completely model independent way as a top-down approach starting with the data to reconstruct the cosmological quantities step by step from the luminosity distance $d_L(z)$ to Hubble parameter $h(z)$ and the deceleration parameter $q(z)$. This method has been used broadly since its proposition and has passed different tests and has shown its strength. Nevertheless, while smoothing method is a promising approach to reconstruct the underlying expansion history of the universe and probably the best way to find a feature in a dataset, estimation of the error-bars on the reconstructed quantities have been an issue since it is not possible to define degrees of freedom in this approach. Error amplification because of taking the derivatives (moving from smooth $d_L(z)$ to $h(z)$ and then $q(z)$) and at the same time controlling the bias in the reconstruction are another important related issues in the smoothing method. In this paper we use the idea of Bayesian interpretation of Crossing Statistic~\cite{Crossing,Crossing_B} to estimate the uncertainties of the reconstructed cosmological quantities in a well defined and robust statistical manner. Bayesian interpretation of Crossing Statistic~\cite{Crossing_B} was introduced recently for the purpose of model selection and falsifying cosmological models in a purely model independent manner while we have no information about the underlying model of the universe. In this paper by combining the two methods, error-sensitive smoothing method and the Crossing Statistic, we reconstruct the expansion history of the universe and estimate the redshift evolution of some cosmological quantities such as deceleration parameter $q(z)$ and Om diagnostic $Om(z)$.


In the following we will explain first briefly about the smoothing method and how it can be used for the reconstruction of the properties of dark energy. Next we explain about the Crossing Statistic and its Bayesian interpretation and then we introduce how we can combine the two methods to reconstruct the expansion history of the universe and set the confidence limits. Efficiency of the method will be tested using simulated data and it will be shown how precisely it can reconstruct the expansion history of the universe. In fact it is not unfair to claim that the proposed method works with considerably higher precision in comparison to other available non-parametric methods while here we are not even using any priors in the analysis. Then I apply the method on the current supernovae and Baryon Acoustic Oscillation (BAO) data. At the end I summarize and highlight the advantages of this approach to other available methods.


\section{Method and Analysis}                        
\label{meth}

\subsection{Smoothing Method}

Smoothing method is a completely model independent approach to derive the $d_L(z)$ relation directly from the data, without any assumptions other than the introduction of a smoothing scale. The only parameter used in the smoothing method is the smoothing width $\Delta$, which is constrained only by the quality and quantity of the data, and has nothing to do with any cosmological model. The smoothing method is an iterative procedure with each iteration typically giving a better fit to the data. It has been shown in \cite{Shafieloo06,Shafieloo07,Shafieloo10} that the final reconstructed results are independent of the assumed initial guess, $d_L(z_i)^g$ below. 

The modified smoothing method (error-sensitive) can be summarized by the following equation~\cite{Shafieloo10}:\\

\begin{eqnarray}
\label{eq:bg}
&&\ln d_L(z,\Delta)^{\rm s}=\ln
\ d_L(z)^g \nonumber\\
&& +N(z) \sum_i \frac{\left [ \ln d_L(z_i)- \ln
\ d_L(z_i)^g \right]}{\sigma^2_{d_L(z_i)}} 
\ {\rm exp} \left [- \frac{\ln^2 \left
( \frac{1+z_i}{1+z} \right ) }{2 \Delta^2} \right ],  \nonumber \\
&&N(z)^{-1}=\sum_i {\rm exp} \left
[- \frac{\ln^2 \left ( \frac{1+z_i}{1+z} \right ) }{2 \Delta^2} \right ] \frac{1}{\sigma^2_{d_L(z_i)}} ~. 
\end{eqnarray}

where $d_L(z)$ is the data, $N(z)$ is the normalization factor, $d_L(z_i)^g$ is the initial guess model and $\Delta$ is the width of smoothing. 

The absolute brightness of the supernovae is degenerate with $H_0$ since the observed quantity is the distance modulus $\mu(z)$. The outcome of the smoothing method is therefore $H_0d_L(z)/c\equiv d_L^{rec}(z)=(1+z)D(z)$. In this paper I choose $\Delta=0.30$ which is similar to the value being used in~\cite{Shafieloo10}. Complete explanation of the relations between the $\Delta$, the number of data points, quality of the data and the reconstructed results can be found in~\cite{Shafieloo06,Shafieloo07}. It has been shown before that smoothing method is a promising approach to reconstruct the expansion history of the universe however, setting the confidence limits have been an issue and previously  bootstrap approach was being used to set the confidence limits. In this paper, the reconstructed form of the $d_L(z)$ will be used as a mean function in the full reconstruction process which includes the idea of Bayesian interpretation of Crossing Statistic as it will be explained in the next section.
 
\subsection{Crossing Statistic and Reconstructing Dark Energy}

The main idea behind this work lies on this fact that the actual model of the universe and the reconstructed models using smoothing method would have one or two mild crossings (the distance modules of the actual model of the universe and the reconstructed $\mu(z)$ from smoothing method would cross each other at one or two points in the data redshift range) and Crossing functions multiplied by the mean functions generated by the smoothing method would cover the actual model of the universe to an indistinguishable level. It is in fact combining a non-parametric method with a parametric method to define and set the confidence limits. Crossing function is defined by Chebichev polynomials~\cite{Crossing_B}:

\begin{equation}
T_{II}(C_1,C_2,z)=1+C_1(\frac{z}{z_{max}})+C_2[2(\frac{z}{z_{max}})^2-1],
\end{equation}

and we fit $\mu_{Smooth}^{T_{II}}(z)=\mu_{Smooth}(z) \times T_{II}(C_1,C_2,z)$ to the data and find the best fit point $C^{best}_1,C^{best}_2$ in the hyperparameter space and also $C_1,C_2$ points related to the $1\sigma$, $2\sigma$ and $3\sigma$ confident limits. Each  $T_{II}(C_1,C_2,z)$ (where $C_1,C_2$ pairs are within hyperparameter confidence contours) multiplied to $\mu_{Smooth}(z)$ represents a reconstruction of the expansion history of the universe consistent to the data.

We should note that though the reconstructed luminosity distances are model independent, the derived uncertainties do depend on the number of Chebichev terms included. In fact depends on the number of Chebichev terms, the $\Delta \chi^2$ (with respect to the best fit point) for different confidence limits can vary and this can affect the width of the error-band. In this paper to estimate the errors we have included the Chebichev polynomials up to the second order and this selection is based on an argument we have discussed in~\cite{Crossing_B}. The fact is that within the range of the data up to redshift of around $z=2$, we cannot expect to have more than two crossings between different assumed models. This is even valid for models with significant differences, e.g  SCDM and LCDM models that though they are so different in their predictions for the expansion history of the universe they do not cross each other in more than two points (for different values of matter density). This is due to the fact that the observables, $\mu(z)$ are integrated quantities over the inverse of Hubble parameter (expansion history) and they monotonically increases by redshift. In~\cite{Crossing_B} we have discussed this issue in more details that the data is not sensitive to more than two crossing between models (equivalently we can include up to the second order of Chebichev polynomials for error estimation). However, this is true that in some cases even assuming one crossing term (using Chebichev polynomial of the first term only) would be sufficient for the purpose of error estimation but it is always safer to consider all unexpected possibilities. In other words, having a broader error-bands which contains the fiducial model with a very high certainty would be more reliable and acceptable than having a very narrow error-band for the reconstructed results but with possibilities of missing the fiducial model at some redshift ranges. As we will see in the results, in all assumed cases for different models of dark energy (even for the non trivial kink model) the reconstruction method works pretty well and the fiducial models are covered by the error-bands at all redshift ranges.

To test the method we apply first the procedure to the simulated data. I assume three sets of simulated data based on three dark energy models (all spatially flat universes). First model is an evolving dark energy model or Kink model with the same parameters used in~\cite{LANL,Crossing_B}. Equation of state of dark energy in this particular model is given by: 

\begin{eqnarray}
\label{eq:model3}
&&w(z) =
w_0+(w_m-w_0)\frac{1+\exp(\Delta_t^{-1}(1+z_t)^{-1})}{1-\exp(\Delta_t^{-1})}\\ 
&&\times\left[ 1-
\frac
{\exp(\Delta_t^{-1}) + \exp(\Delta_t^{-1}(1+z_t)^{-1})}
{\exp(\Delta_t^{-1}(1+z)^{-1}) +\exp(\Delta_t^{-1}(1+z_t)^{-1})} 
\right],\nonumber
\end{eqnarray}
with the constants having the values $w_0 = -1.0, ~w_m = -0.5,~z_t =
0.5,~\Delta_t = 0.05$. I assume $\Omega_m=0.27$ in the simulations. Second model is based on a dark energy model with constant equation of state of $w(z)=0.9$ and with $\Omega_m=0.27$ and finally the third simulated data is based on $\Lambda$CDM model with $w(z)=-1$ and $\Omega_m=0.27$. Simulated data are based on future space based supernovae data with 2298 data points in the range of $0.015<z<1.7$ with intrinsic dispersion of $\sigma_{int}=0.13$~\cite{JDEM}. The Kink model I used in this analysis has a special form of the equation of state of dark energy that at low redshifts it converges to $w(z)=-1$ and at higher redshifts it smoothly changes to $w=-0.5$. This evolving equation of state of dark energy is not reducible to the common form of $w(z)=w_0+w_az/(1+z)$~\cite{CPL} or many other dark energy parametrizations. Hence using usual parameterizations results to wrong reconstruction of dark energy equation of state if the data is based on the Kink model (look at upper-right panel of figure.~6 in~\cite{LANL}).

Deriving $d_L(z)$ we can accordingly derive $h(z)$, $Om(z)$ and $q(z)$: 
\begin{eqnarray}
H(z) =\left[\frac{d}{dz}\left(\frac{d_L(z)}{1+z}\right)\right]^{-1}\\
q(z)=(1+z)\frac{H'(z)}{H(z)}-1\\
Om(z)=\frac{h^2(z)-1}{(1+z)^3-1}
\label{eq:h}
\end{eqnarray}

where derivatives are respect to redshift and $h(z)=H(z)/H_0$. To derive all these quantities we do not need to know the value of the matter density and in particular $Om(z)$ can be used directly to falsify the cosmological constant which requires $Om(z)$ should be a constant at all redshifts~\cite{sahni08}.

We should note that the observables $\mu(z)$ that are tightly related to the luminosity distances $d_L(z)$, are integrated properties that are functions of the evolution of the universe and there might be different dark energy models (assuming different matter density and curvature) that result to the same observables. The concept of cosmographic degeneracy does exist and make it almost impossible to distinguish between some certain dark energy models if we have no tight constraints on the curvature and matter density~\cite{arman_eric}. However, in this paper we try to avoid this issue and we work on $h(z)$ and $q(z)$ directly. The reason is that unlike the equation of state of dark energy $w(z)$, these quantities $h(z)$ and $q(z)$ are direct derivatives of the reconstructed distance results and they are independent of the actual value of the matter density. We have also assumed a flat spatial curvature through out the paper.

\begin{figure*}[!t]
\vspace{-1.8in}
\hspace{0.5in}
\includegraphics[scale=0.50, angle=0]{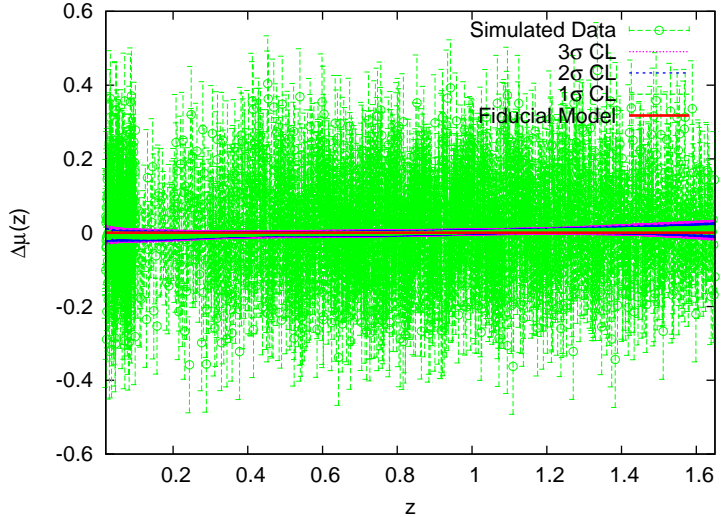}
\vspace{-.7in}
\caption {\small A set of simulated data and the reconstructed $\mu(z)$ and its error bands. The simulated data is based on a spatially flat LCDM model and the actual model is subtracted from the data and the reconstructed results for better clarification.}
\label{fig:mu}
\end{figure*}

To have an idea of the reconstruction process, in figure.~\ref{fig:mu} we show a set of simulated data and the reconstructed $\mu(z)$ and its error bands. The simulated data is based on a spatially flat LCDM model and the fiducial model is subtracted from the data and the reconstructed results for better clarification. In this paper by $1\sigma$, $2\sigma$ and $3\sigma$ we mean $68\%$, $95\%$ and $99\%$ confidence limits respectively. 

\begin{figure*}[!t]
\centering
\begin{center}
\vspace{-0.05in}
\centerline{\mbox{\hspace{0.in} \hspace{2.1in}  \hspace{2.1in} }}
$\begin{array}{@{\hspace{-0.3in}}c@{\hspace{0.3in}}c@{\hspace{0.3in}}c}
\multicolumn{1}{l}{\mbox{}} &
\multicolumn{1}{l}{\mbox{}} \\ [-2.8cm]
\hspace{-0.2in}
\includegraphics[scale=0.20, angle=0]{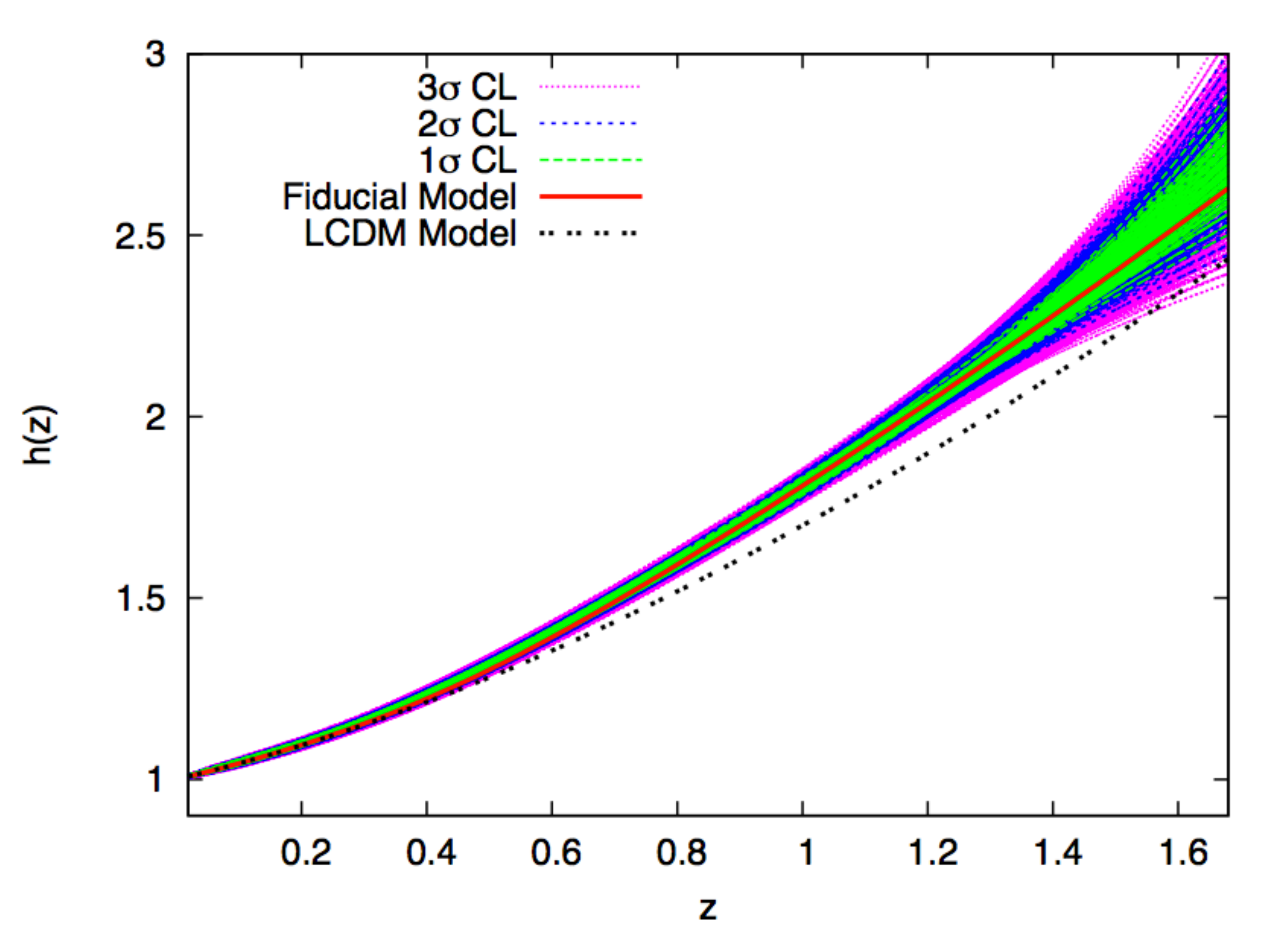}
\hspace{-.in}
\includegraphics[scale=0.20, angle=0]{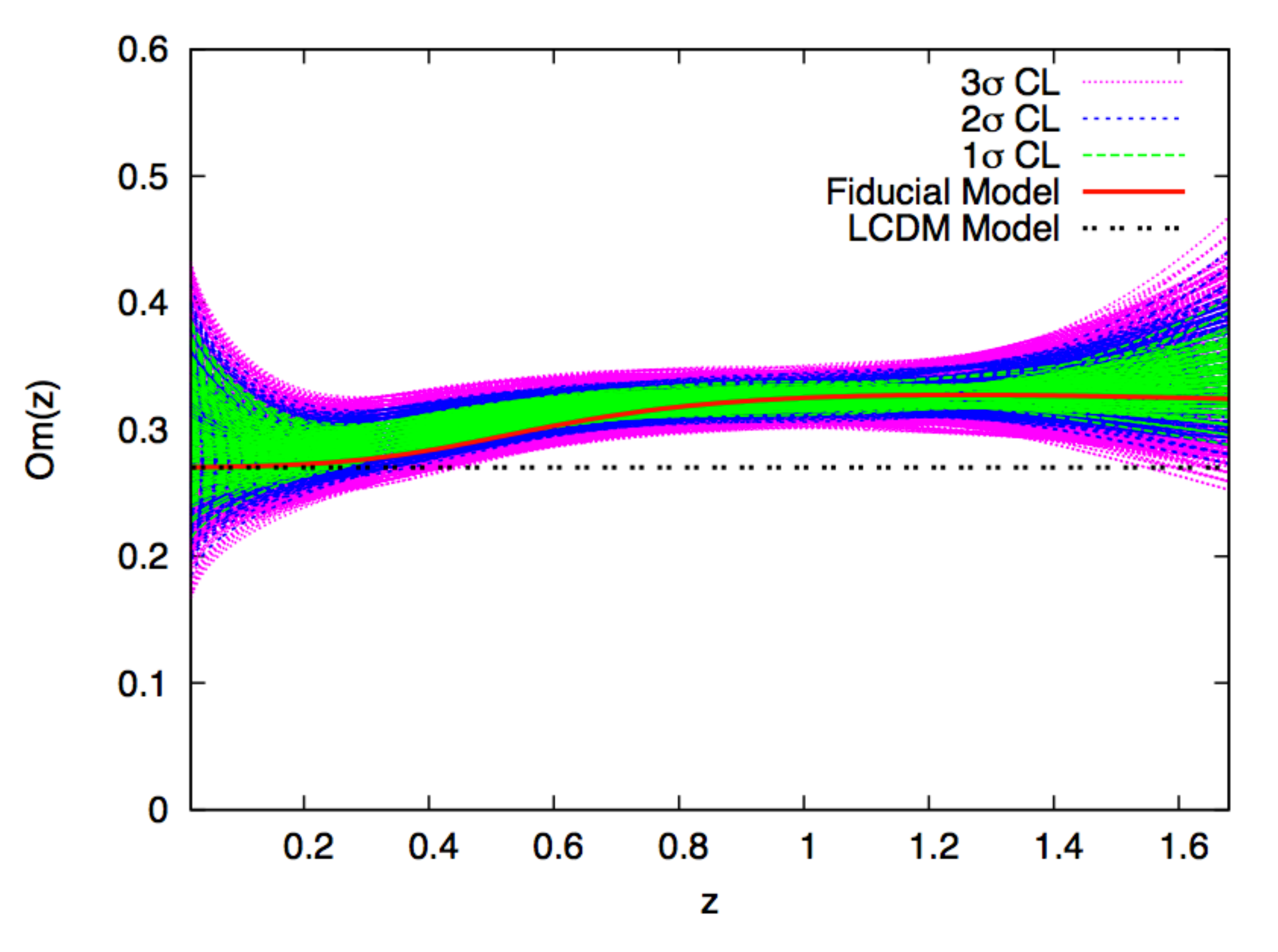}
\hspace{-.in}
\includegraphics[scale=0.20, angle=0]{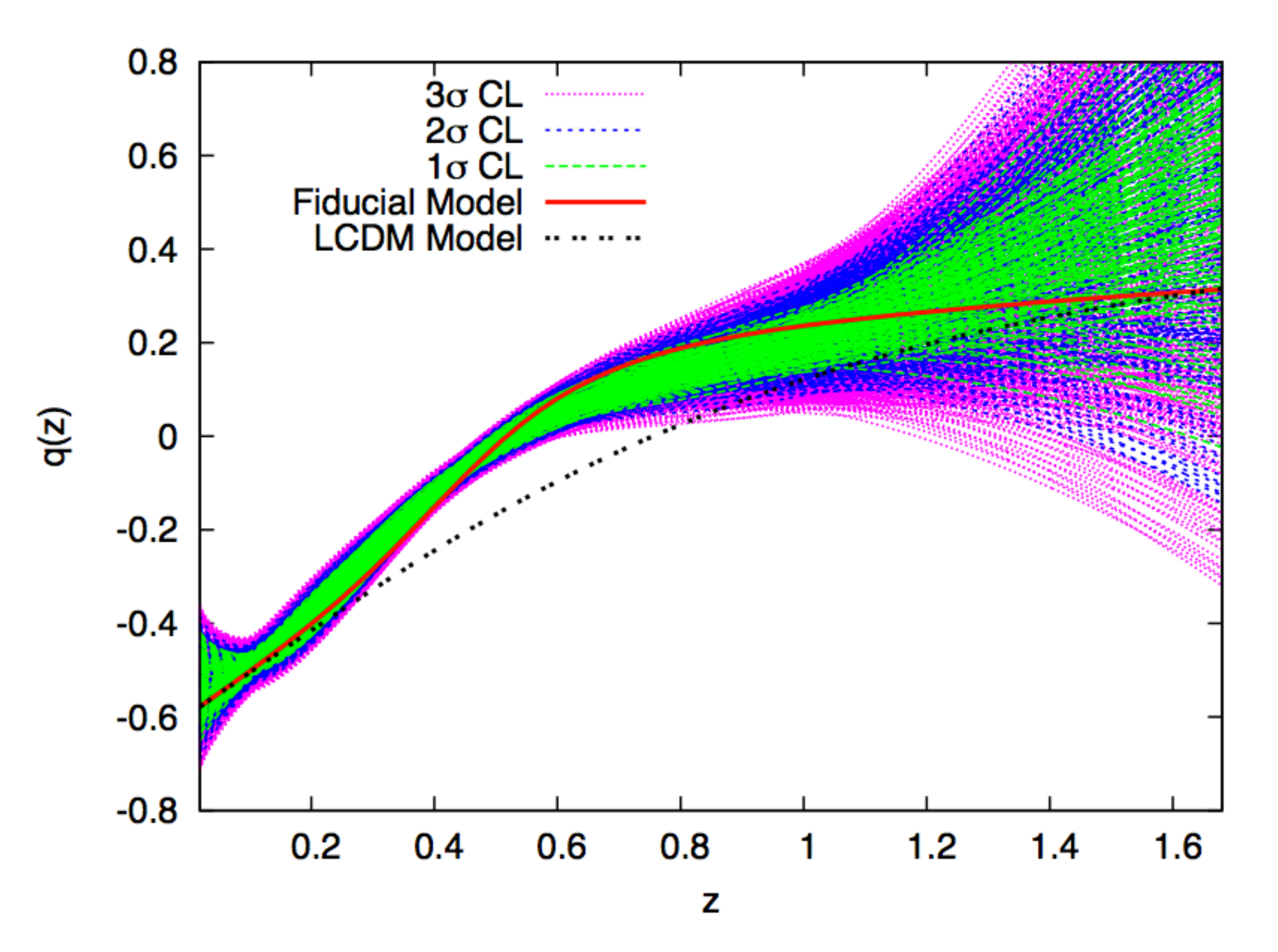}
\hspace{-.1in}
\vspace{0.8in}
\end{array}$
$\begin{array}{@{\hspace{-0.3in}}c@{\hspace{0.3in}}c@{\hspace{0.3in}}c}
\multicolumn{1}{l}{\mbox{}} &
\multicolumn{1}{l}{\mbox{}} \\ [-2.8cm]
\hspace{-0.2in}
\includegraphics[scale=0.20, angle=0]{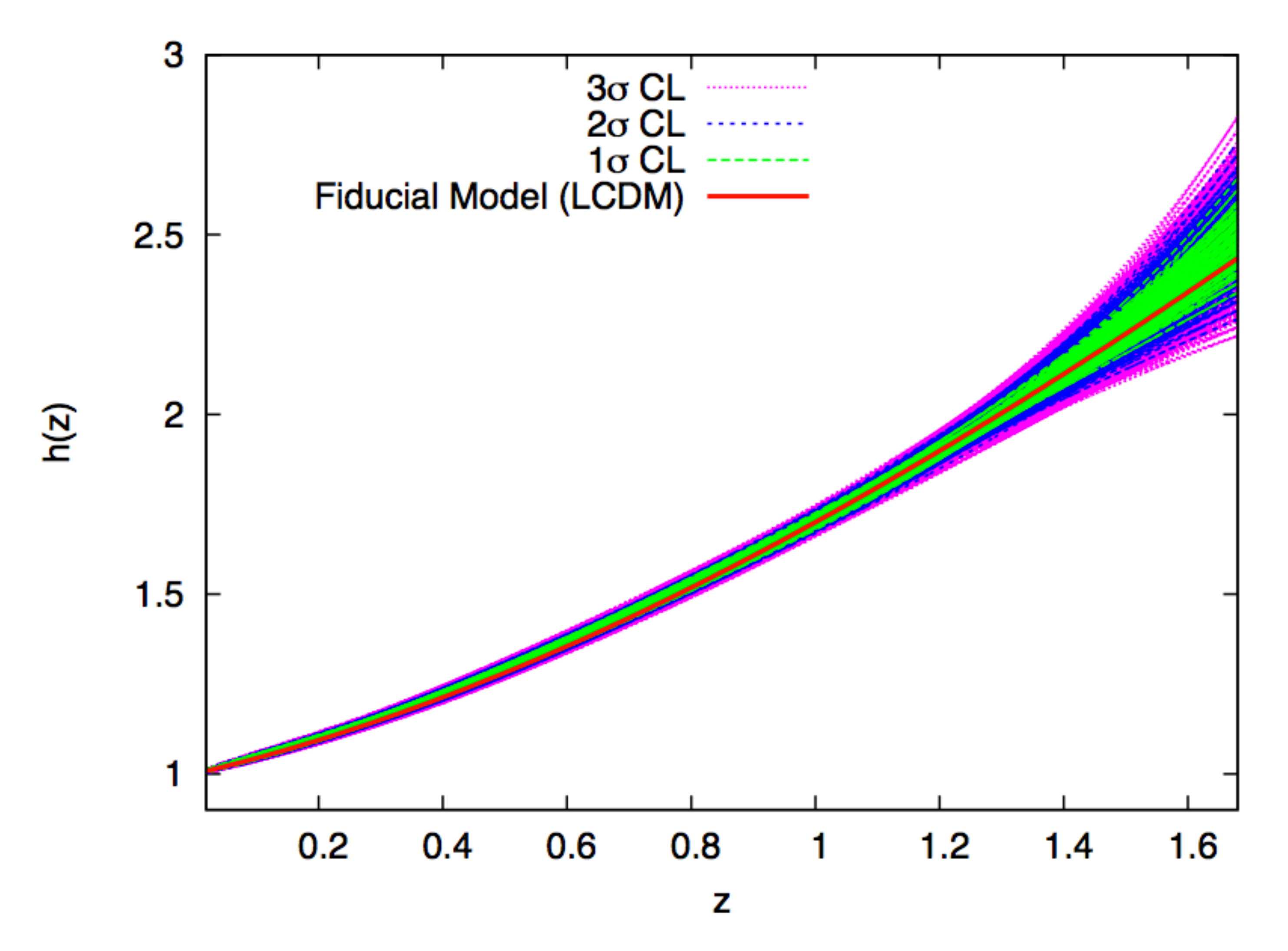}
\hspace{-.in}
\includegraphics[scale=0.20, angle=0]{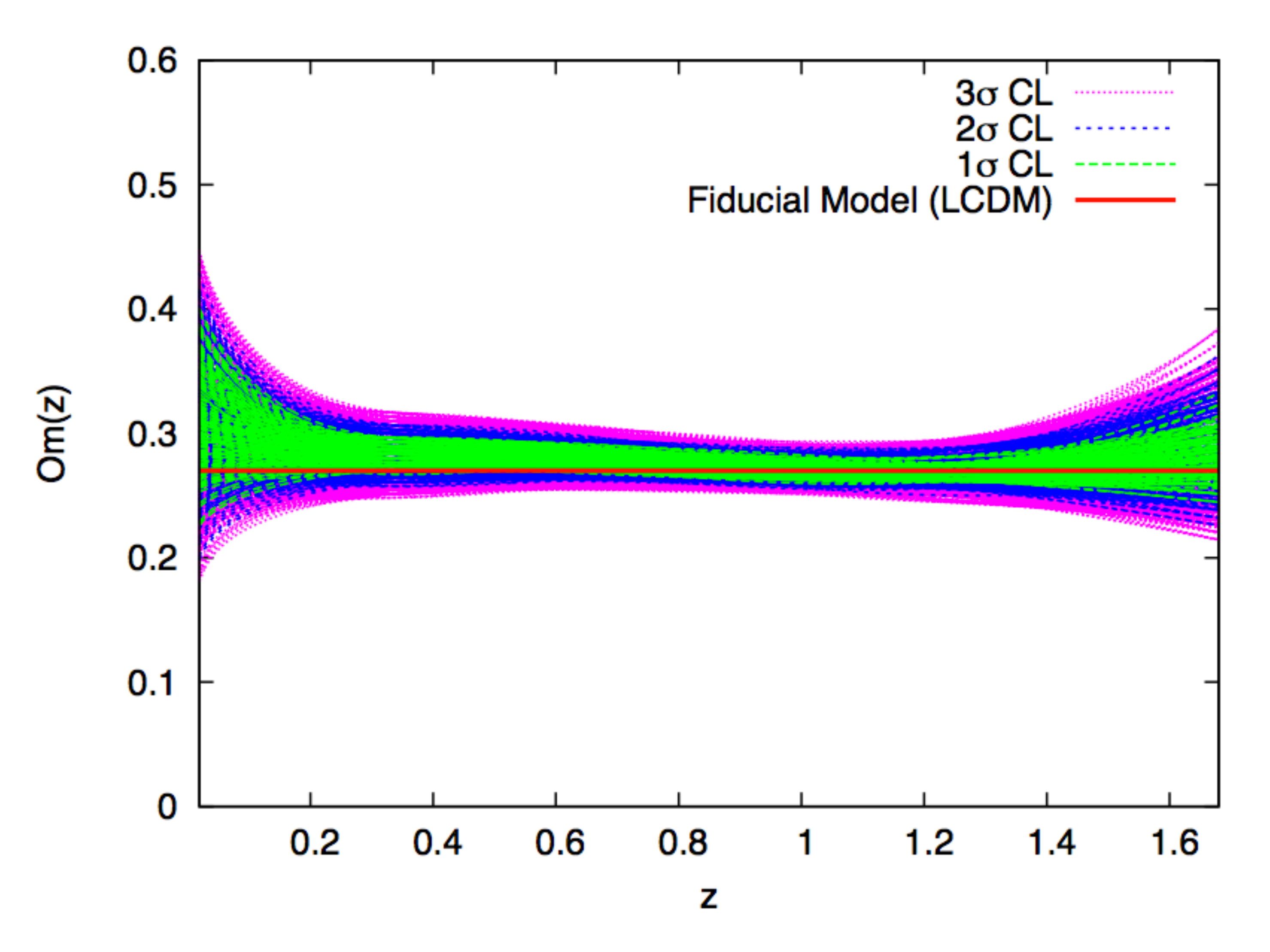}
\hspace{-.in}
\includegraphics[scale=0.20, angle=0]{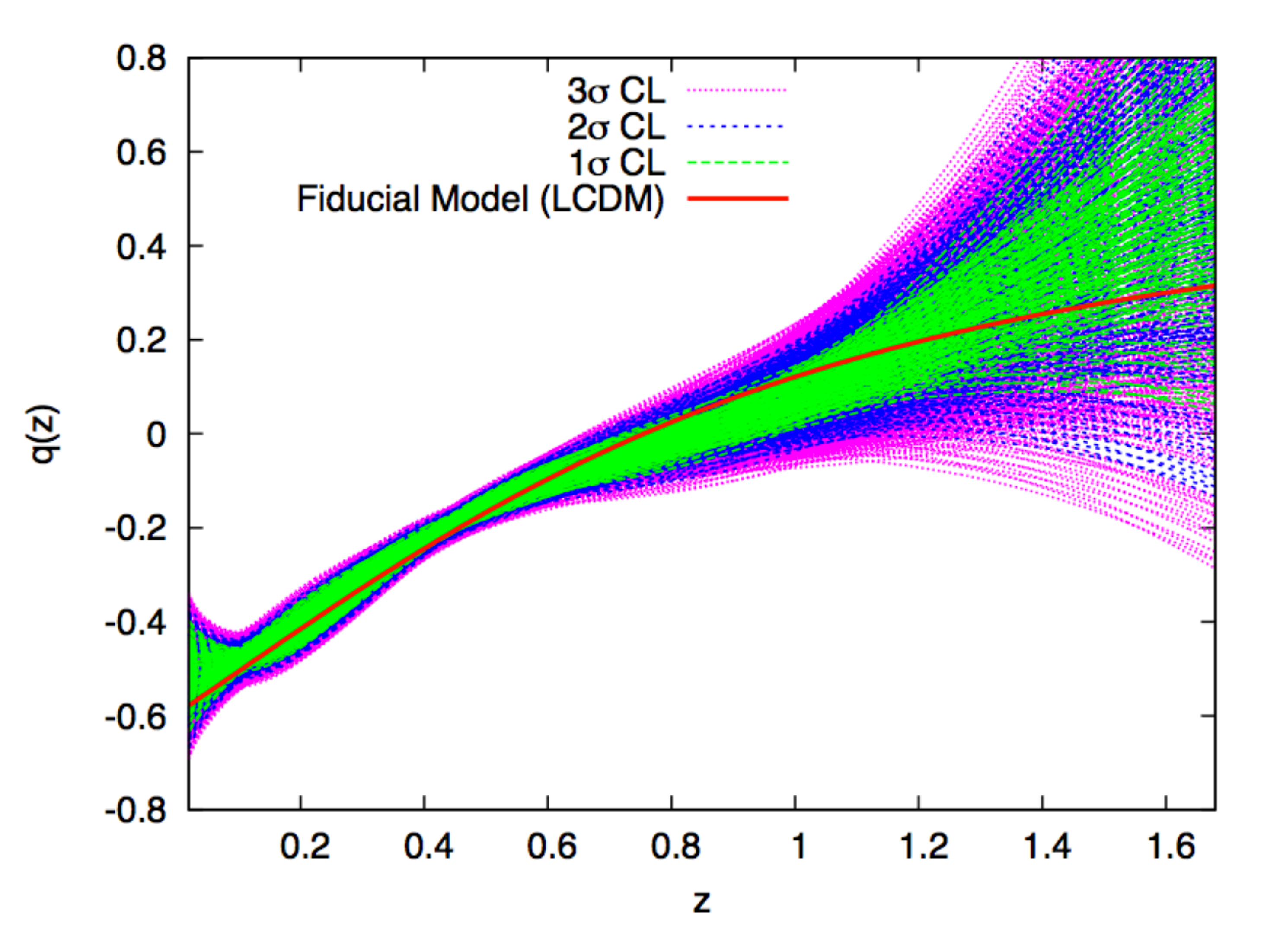}
\hspace{-.1in}
\vspace{0.8in}
\end{array}$
$\begin{array}{@{\hspace{-0.3in}}c@{\hspace{0.3in}}c@{\hspace{0.3in}}c}
\multicolumn{1}{l}{\mbox{}} &
\multicolumn{1}{l}{\mbox{}} \\ [-2.8cm]
\hspace{-0.2in}
\includegraphics[scale=0.20, angle=0]{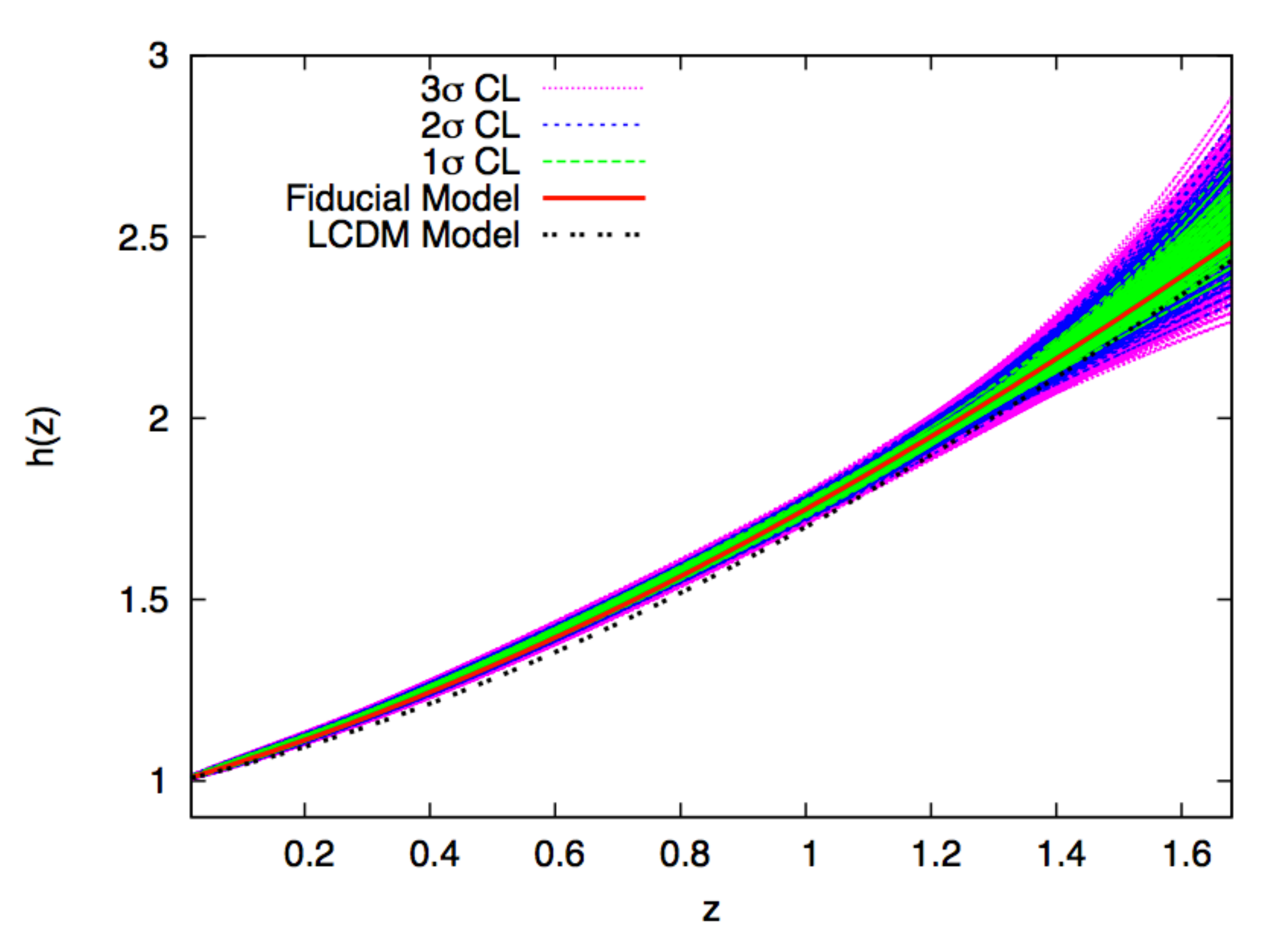}
\hspace{-.in}
\includegraphics[scale=0.20, angle=0]{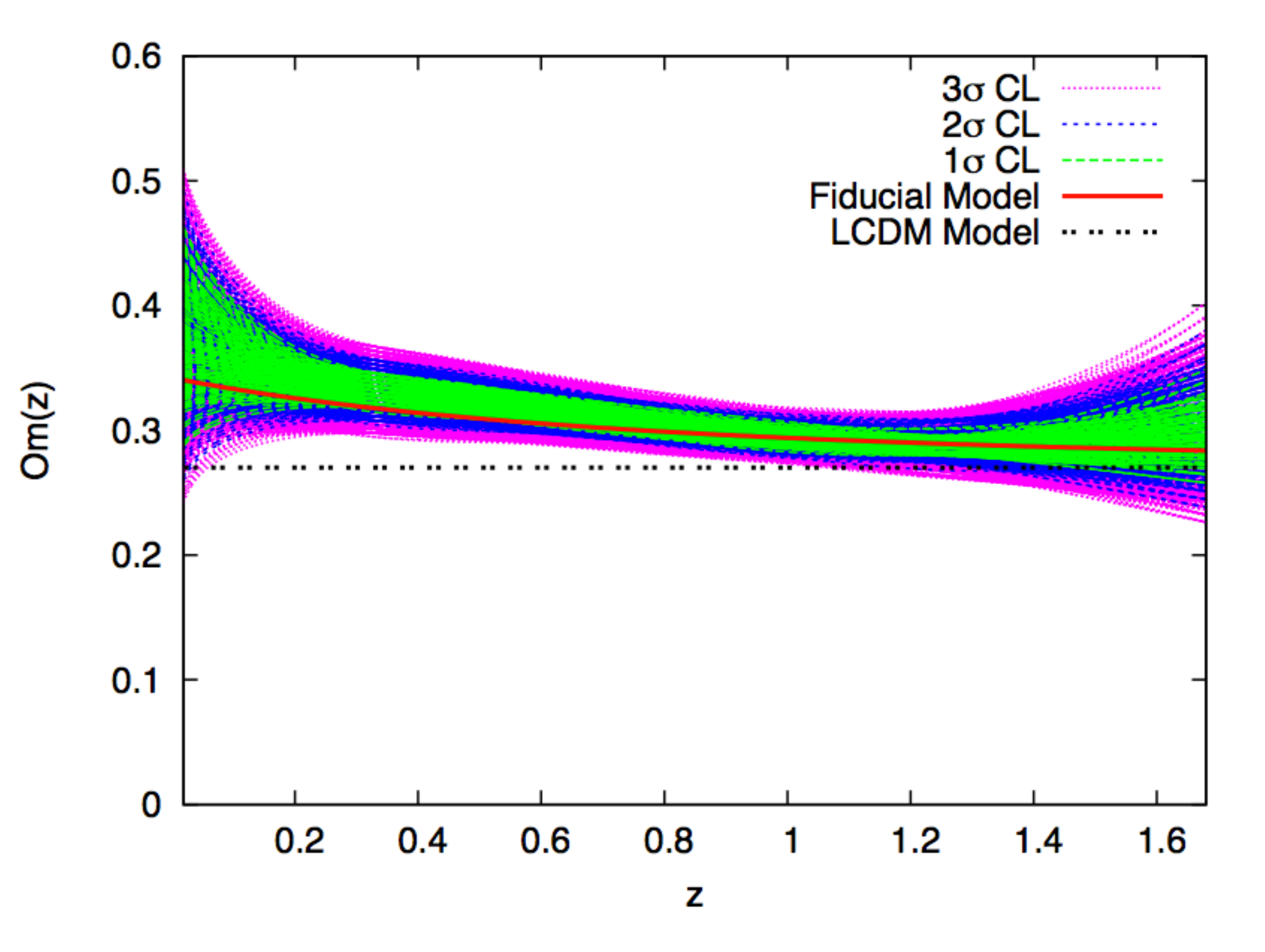}
\hspace{-.in}
\includegraphics[scale=0.20, angle=0]{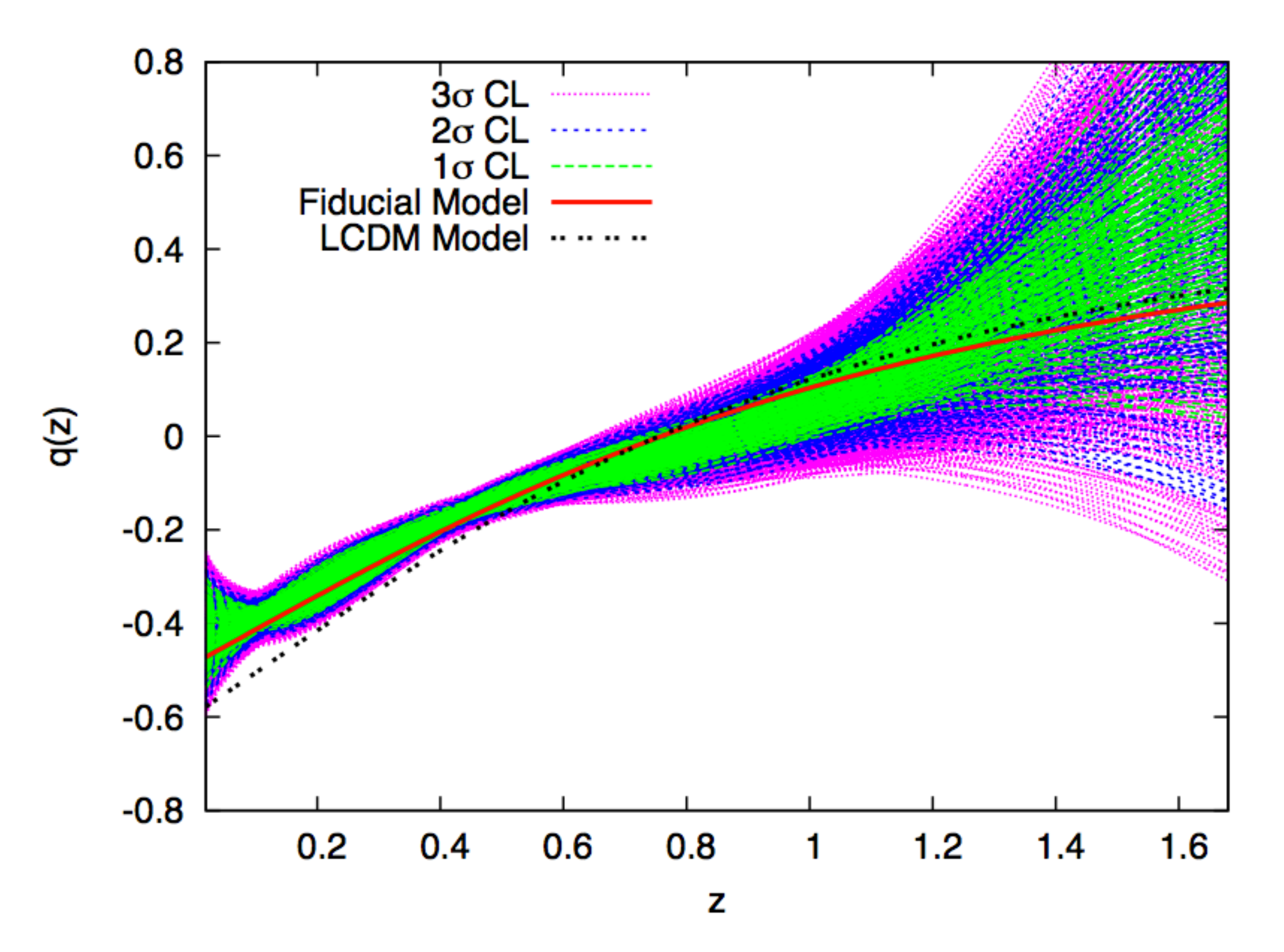}
\hspace{-.1in}
\vspace{-.2in}
\end{array}$
\end{center}
\caption {\small Reconstructed $h(z)$, $Om(z)$ and $q(z)$ with their $1\sigma$ (green), $2\sigma$ (blue) and $3\sigma$ (magenta) confidence limits for three sets of the simulated JDEM data based on a kink dark energy model on the top, $\Lambda$CDM in the middle and $w=-0.9$ in the bottom, all with $\Omega_{0m}=0.27$. Strength of the method in reconstructing the cosmological quantities is obvious. Black double-dot lines represents spatially flat $\Lambda$CDM model with $\Omega_{0m}=0.27$ for comparison.}
\label{fig:sim}
\end{figure*}

In figure.~\ref{fig:sim} we see the reconstructed $h(z)$, $Om(z)$ and $q(z)$ for the three assumed models. One can clearly see that the reconstruction procedure performs perfectly well in reconstructing the cosmological quantities. One should notice that here we are not assuming any particular prior on any cosmological quantity which gives it an advantage to other reconstruction processes.

\subsection{Real Current SN Ia and BAO Data}

Now we can apply the method on the current available data. We use Union2.1 supernovae compilation~\cite{Union21} that is the most recent compilation of the supernovae data along with WiggleZ baryon acoustic oscillation data~\cite{WiggleZ}. Using BAO data we consider the $d_z$ ratios all scaled to $d_z$ at $z=0.106$ to make the BAO data independent of the knowledge of the early universe (look at table.~\ref{table:BAOdata}). $d_z$ is given by:

\begin{equation}
d_z=\frac{r_s(z_{CMB})}{D_V(z)}
\end{equation}

where $r_s(z_{CMB})$ is the sound horizon at the epoch when CMB photons decouple from baryons and $D_V(z)$ is the dilation-scale distance,

\begin{equation}
D_V(z)=\left[\frac{d_L(z)^2}{(1+z)^2}\left(\frac{cz}{H(z)}\right)\right]^{1/3}\\
\end{equation}

\begin{figure*}[!t]
\centering
\begin{center}
\vspace{-0.05in}
\centerline{\mbox{\hspace{0.in} \hspace{2.1in}  \hspace{2.1in} }}
$\begin{array}{@{\hspace{-0.3in}}c@{\hspace{0.3in}}c@{\hspace{0.3in}}c}
\multicolumn{1}{l}{\mbox{}} &
\multicolumn{1}{l}{\mbox{}} \\ [-2.8cm]
\hspace{0.2in}
\includegraphics[scale=0.25, angle=0]{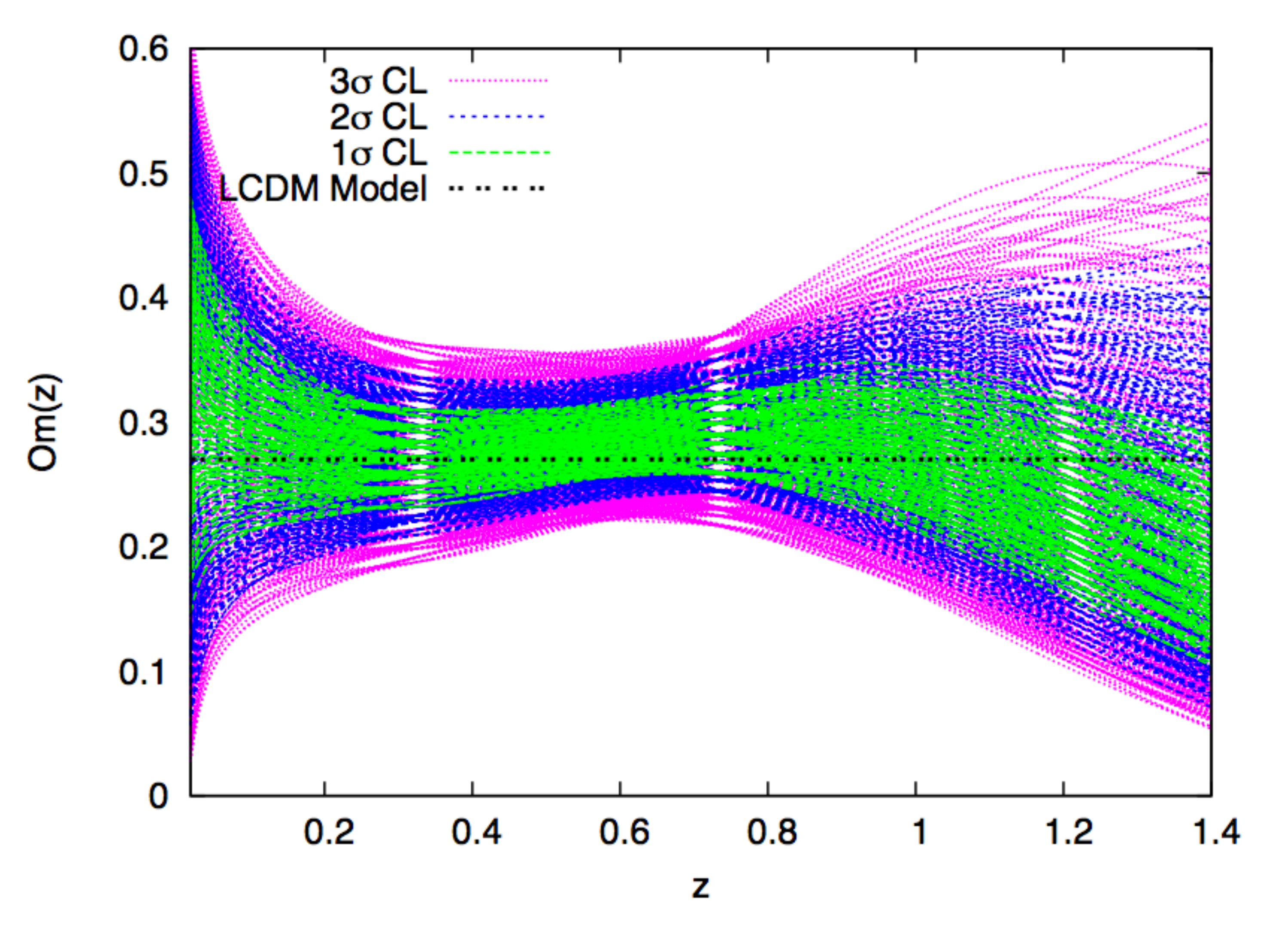}
\hspace{.2in}
\includegraphics[scale=0.25, angle=0]{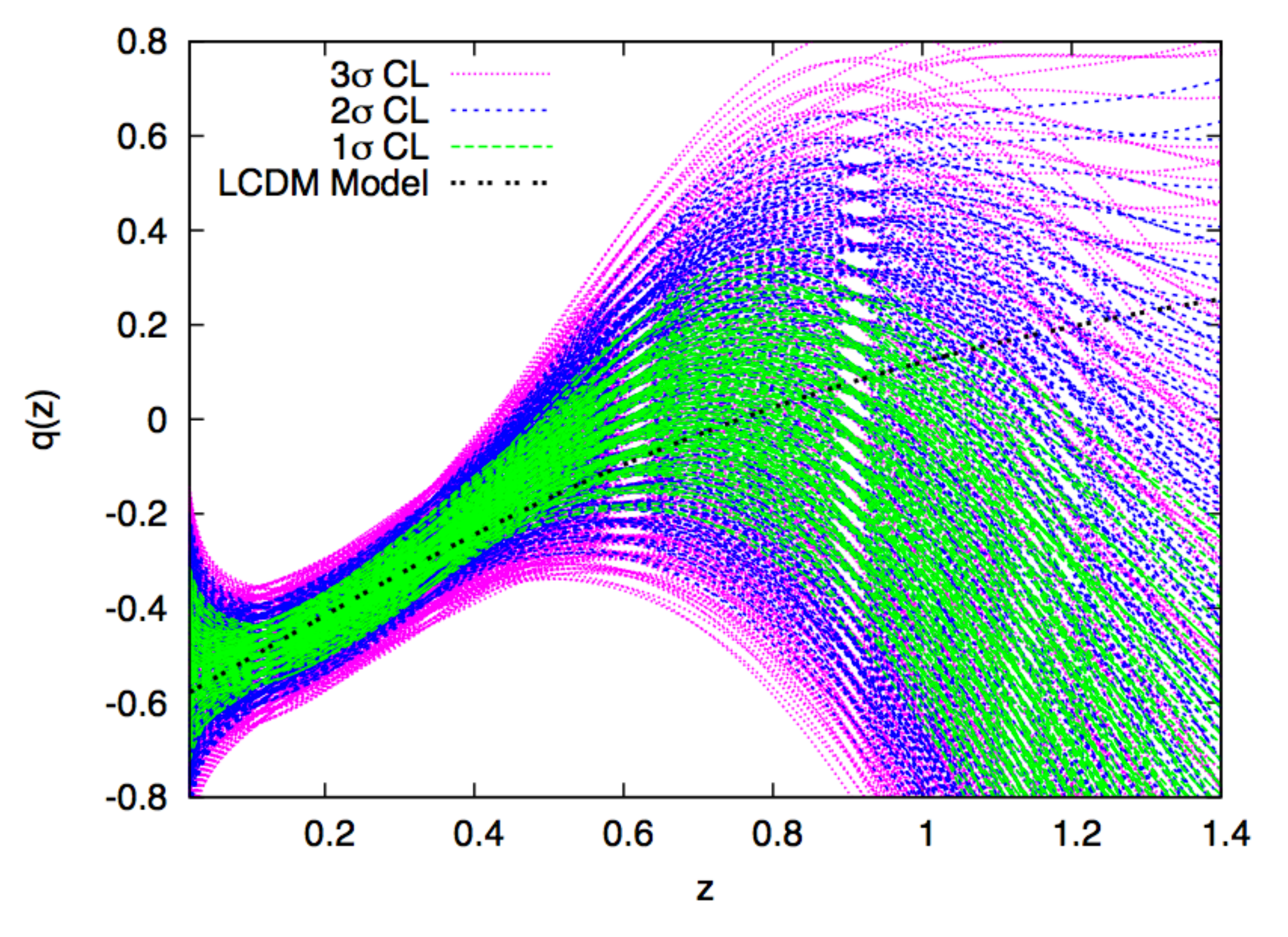}
\hspace{-.1in}
\vspace{0.8in}
\end{array}$
$\begin{array}{@{\hspace{-0.3in}}c@{\hspace{0.3in}}c@{\hspace{0.3in}}c}
\multicolumn{1}{l}{\mbox{}} &
\multicolumn{1}{l}{\mbox{}} \\ [-2.8cm]
\hspace{0.2in}
\includegraphics[scale=0.25, angle=0]{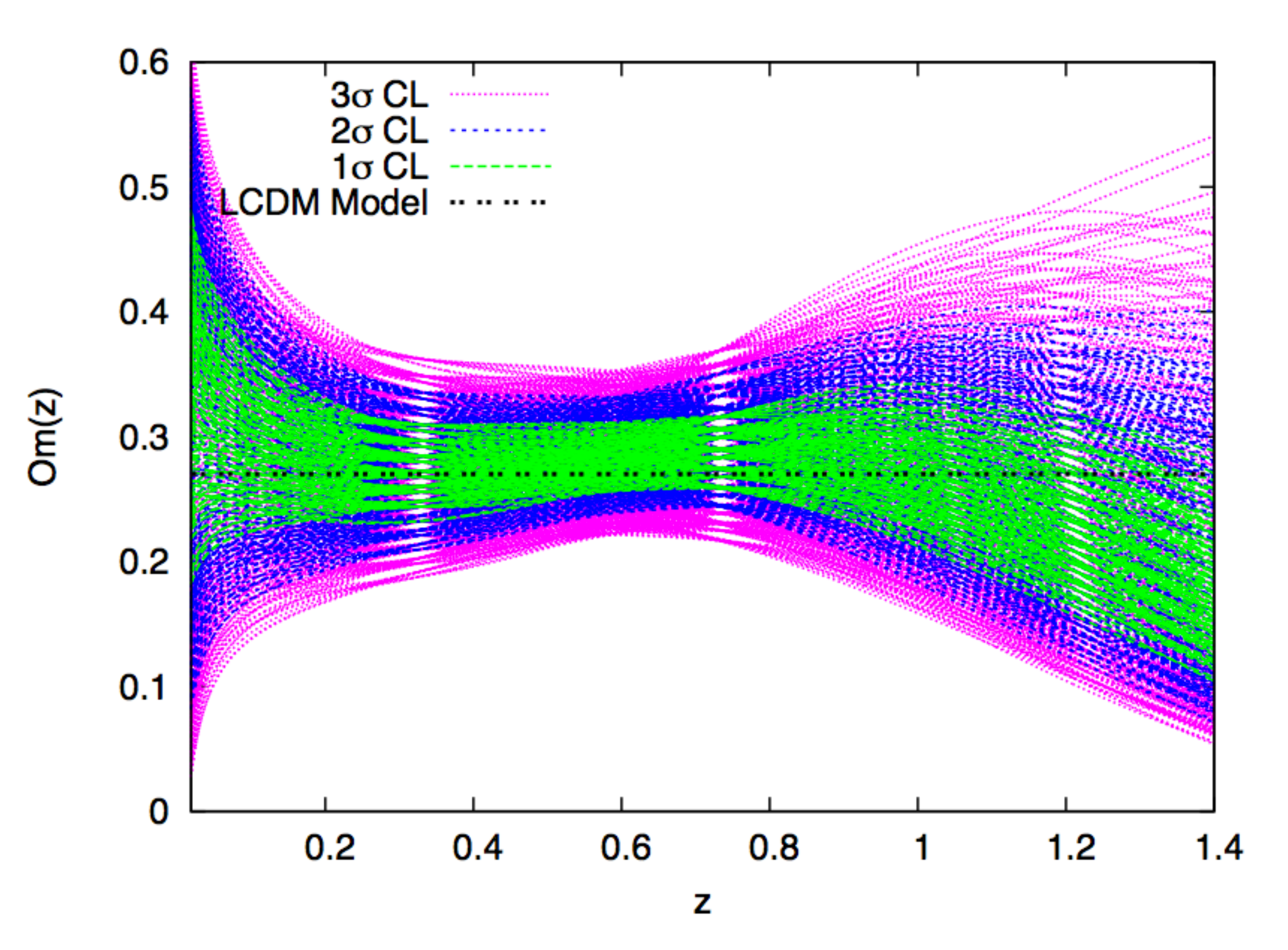}
\hspace{.2in}
\includegraphics[scale=0.25, angle=0]{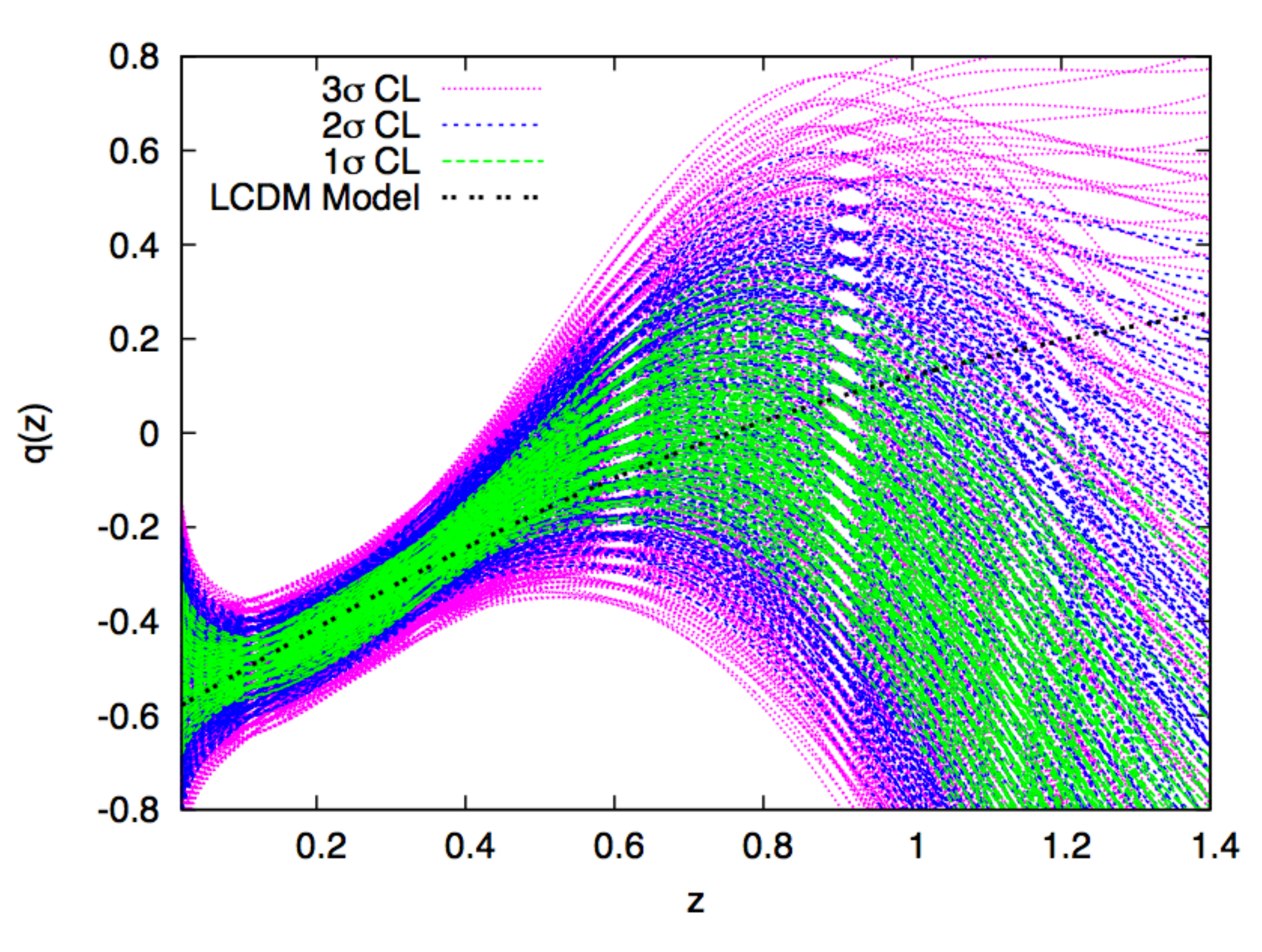}
\hspace{-.1in}
\vspace{-.2in}
\end{array}$
\end{center}
\caption {\small Reconstructed $Om(z)$ and $q(z)$ with their $1\sigma$ (green), $2\sigma$ (blue) and $3\sigma$ (magenta) confidence limits using Union2.1 supernovae data (on the top panels) and using combination of Union2.1 supernovae and WiggleZ BAO data (on the bottom panels). One can see that including BAO data into the analysis does not change the reconstructed results significantly. At the current status of the cosmological observations supernovae data contain much more information on the expansion history of the universe in comparison to BAO data.}
\label{Fig:union21}
\end{figure*}

\begin{figure*}[!t]
\hspace{0.8in}
\includegraphics[scale=0.30, angle=0]{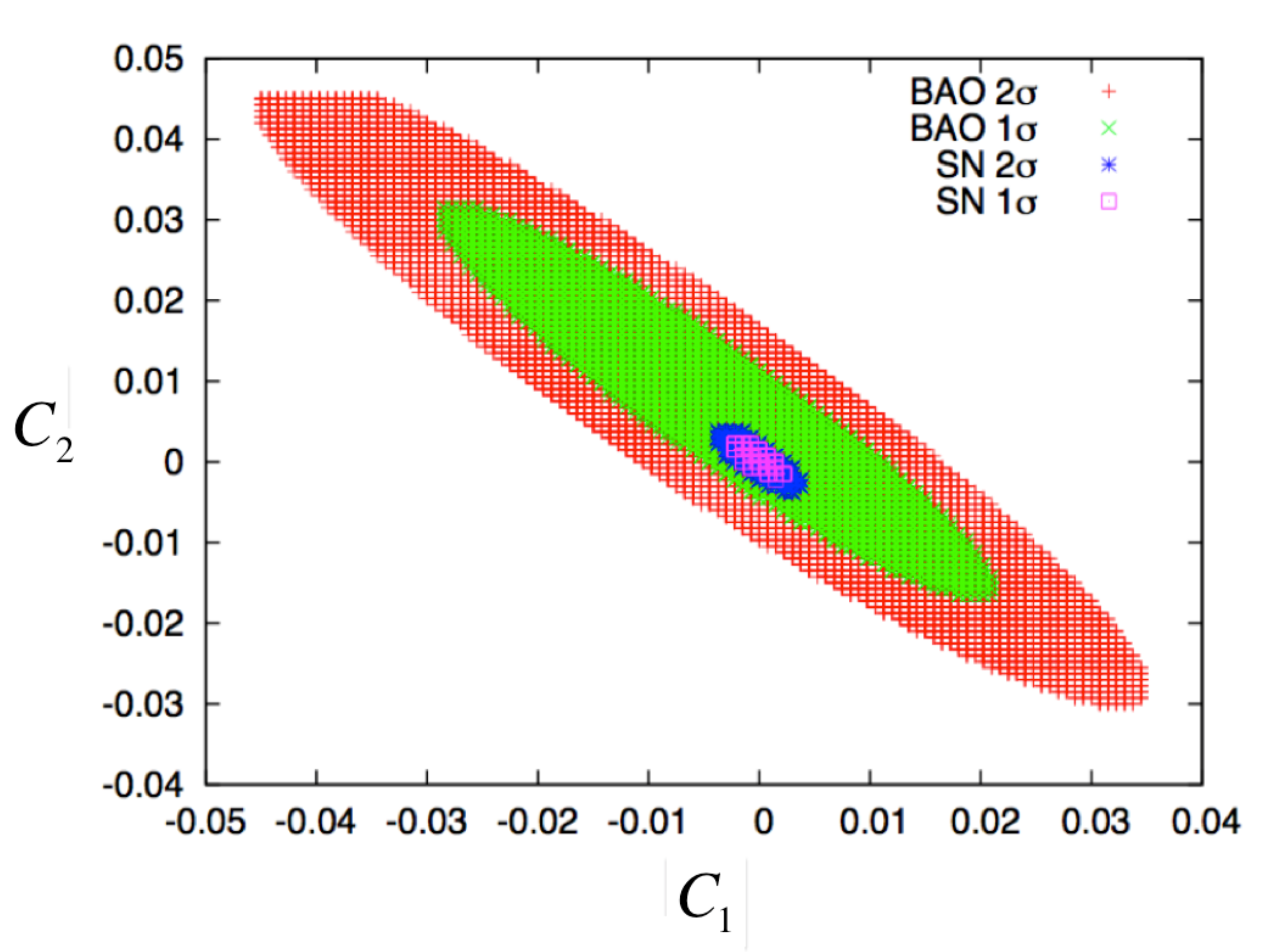}
\caption {\small Confidence contours of the Crossing hyperparameters using Union 2.1 supernovae data ($1\sigma$ CL in magenta, $2\sigma$ CL in blue) and WiggleZ BAO data ($1\sigma$ CL in green, $2\sigma$ CL in red). There is a clear consistency between the two data sets without assuming any particular cosmological model. One can also see that supernovae data are more powerful in putting constraints on the expansion history of the universe in comparison to BAO data at the current status of cosmological observations.}
\label{Fig:contours}
\end{figure*}

In fact by using the $d_z$ ratios we remove the $r_S(z_{CMB})$ from the equations and minimize the effects of the assumptions of the early universe from our analysis. Similar to the procedure we used in the previous section we apply first the smoothing method on the supernovae data to reconstruct the mean functions and then we use these mean functions along with Crossing functions fitting supernovae and BAO data to find out the Crossing hyper parameters and define the confidence limits. Reconstructed $Om(z)$ and $q(z)$ using Union 2.1 data is shown in figure.~\ref{Fig:union21} upper panels. In the lower panels we can see the same quantities using combination of Union 2.1 supernovae and WiggleZ BAO data. One can clearly see that the standard flat $\Lambda$CDM model has a proper consistency to the data. In figure.~\ref{Fig:contours} we see the $1\sigma$ and $2\sigma$ confidence contours of $C_1, C_2$ hyperparameters using supernovae data and BAO data individually. One should remember that these are the coefficients of the Crossing function around the mean function generated by smoothing method using supernovae data. So it is expected that the $C_1,C_2$ contours using supernovae data be centered around $0,0$ point. This plot shows that the BAO and supernovae data are in proper consistency with each other without assuming any cosmological model. It also shows that supernovae data is still a much more powerful probe of the expansion history of the universe in comparison to BAO data (notice the size of the confidence contours). One reason lies on the fact that supernovae data covers a very large redshift range (from as low as $z=0.015$ up to about $z=1.5$) and also the number of data points are considerably larger in the supernovae compilations.

\begin{table*}
\begin{tabular}{ccccccccc}
 $$ &\hspace{12 mm} Inverse Covariance & \hspace{-4 mm} Matrix & & $$\\ 
\hline
 $$ &\hspace{-30 mm} $\frac{d_{z=0.106}}{d_{z=0.20}}$ & \hspace{-55 mm} $\frac{d_{z=0.106}}{d_{z=0.35}}$ & \hspace{-35 mm} $\frac{d_{z=0.106}}{d_{z=0.44}}$ & \hspace{-10 mm}$\frac{d_{z=0.106}}{d_{z=0.60}}$ &  \hspace{1 mm} $\frac{d_{z=0.106}}{d_{z=0.73}}$ &  \hspace{2 mm} Mean Values \vspace{1mm}\\
\hline
$d_{z=0.106}/d_{z=0.20}$& \hspace{-30 mm} $284.709$ & \hspace{-55 mm}$-103.845$ & \hspace{-35 mm}$-4.773$ & \hspace{-10 mm}$-9.600$ & \hspace{1 mm}$-6.961$  &  \hspace{2 mm} $1.764 \pm 0.097$ \\
\hline
$d_{z=0.106}/d_{z=0.35}$& \hspace{-30 mm} $-103.845$ & \hspace{-55 mm}$ 91.940$ & \hspace{-35 mm}$-2.552$  & \hspace{-10 mm}$-5.133$ & \hspace{1 mm}$-3.722$ &  \hspace{2 mm} $ 3.063 \pm 0.170$ \\
\hline
$d_{z=0.106}/d_{z=0.44}$& \hspace{-30 mm} $-4.773$ & \hspace{-55 mm}$-2.552$ & \hspace{-35 mm}$14.546$  & \hspace{-10 mm}$-9.575$ & \hspace{1 mm}$2.275$ &  \hspace{2 mm} $3.668 \pm 0.328$ \\
\hline
$d_{z=0.106}/d_{z=0.60}$& \hspace{-30 mm} $-9.600$ & \hspace{-55 mm}$-5.133$ & \hspace{-35 mm}$-9.575$  & \hspace{-10 mm}$30.340$ & \hspace{1 mm}$-10.771$ &  \hspace{2 mm} $4.628 \pm 0.299$ \\
\hline
$d_{z=0.106}/d_{z=0.73}$& \hspace{-30 mm} $-6.961$ & \hspace{-55 mm}$-3.722$ & \hspace{-35 mm}$2.275$  & \hspace{-10 mm}$-10.771$ & \hspace{1 mm}$12.954$ &  \hspace{2 mm} $ 5.676 \pm 0.398 $ \\
\hline
\end{tabular}
\caption{\small WiggleZ baryon acoustic oscillation distance ratios scaled to $d_{z=0.106}$ and their inverse covariance matrix~\cite{Blake}. Using the BAO distance ratios one can reduce the effects from the early universe on the distance measurements. }
\label{table:BAOdata}
\end{table*}

\section{Conclusion}                        
\label{concl}

In this paper we present a robust and easy to use method of model independent reconstruction of the expansion history of the universe which performs perfectly well for any dark energy model. Combining the smoothing method and Crossings Statistic results in an approach which can be easily used to reconstruct the expansion history of the universe and the properties of dark energy without setting any prior on the cosmological quantities. This is an important advantage over other available reconstruction methods in which one has to set some priors on the equation of state of dark energy or some other cosmological parameters. The presented approach also resolves the problem of defining the confidence limits around the reconstructed cosmological quantities using smoothing method. It has been shown before that smoothing method is a very powerful approach to reconstruct the expansion history of the universe but because of its non-parametric nature, it was not possible to define the confidence limits in a straightforward way. In this work, using the idea of Bayesian interpretation of Crossing Statistic this problem is resolved and one can set proper confidence limits on the reconstructed quantities. We have shown the strength of the method in reconstruction of the expansion history of the universe using simulated data where we could reconstruct different dark energy models, even some non trivial ones, with a very high precision. We have also used the method to test the consistency of the two important cosmological data sets, Union 2.1 supernovae data and WiggleZ BAO data in a completely model independent way. These two data sets are pretty well consistent with each other, however, at the current status of cosmological observations supernovae data has much more power in reconstruction of the expansion history of the universe. Based on our analysis spatially flat $\Lambda$CDM model has a proper concordance to the given Union 2.1 supernovae and WiggleZ BAO data, even though, we cannot still rule out many other dark energy models including other two models discussed in this paper.

\acknowledgments{AS thanks Eric Linder and Chris Blake for useful discussions. This work has been supported by World Class University grant R32-2009-000-10130-0 through the National Research Foundation, Ministry of Education, Science and Technology of Korea.}

\appendix

\end{document}